\documentclass{article}
\usepackage[utf8]{inputenc}
\usepackage{graphicx}
\usepackage{float}
\usepackage{geometry}
\usepackage{amsmath}
\usepackage{amssymb}
\usepackage{color}
\usepackage[hidelinks]{hyperref}
\usepackage{microtype}
\usepackage{xcolor}
\usepackage{authblk}

\newcommand{\thi}{\theta_{\mathrm{hi}}}
\newcommand{\tlo}{\theta_{\mathrm{lo}}}

\DeclareRobustCommand{\ind}{\mathchoice
  {\mathrm{1\mskip-4.2mu l}}{\mathrm{1\mskip-4.2mu l}}%
  {\mathrm{1\mskip-4.6mu l}}{\mathrm{1\mskip-5.2mu l}}}

\title{Let Them Peek: Group Sequential Designs for A/B Testing}
\author{Lorenzo Perini}
\author{Mia Garrard}
\author{Rina Friedberg}
\author{Rohit Kumar}
\author{Houssam Nassif}
\author{Abbas Zaidi}
\author{Pablo Alcain}
\affil{Meta Platforms, Inc.}
\date{}

\let\OLDthebibliography\thebibliography
\renewcommand\thebibliography[1]{
  \OLDthebibliography{#1}
  \setlength{\parskip}{0pt}
  \setlength{\itemsep}{0pt plus 0.3ex}
}

\begin{document}

\setlength{\abovedisplayskip}{4pt}
\setlength{\belowdisplayskip}{4pt}
\setlength{\abovedisplayshortskip}{0pt}
\setlength{\belowdisplayshortskip}{4pt}

\maketitle

\section{Introduction}
\label{sec:intro}

Inferring causal effects of treatments is a goal in the evaluation of product innovations, and randomized controlled trials, commonly known as A/B tests, are the standard mechanism for doing so~\cite{kohavi2020trustworthy}. Under the potential outcomes framework, the causal estimand is a comparison of the potential outcomes of the same units under different treatment conditions; on an experimentation platform it is summarized by a cumulative lift estimated from the accumulating data. Standard platforms evaluate these tests using fixed-horizon procedures: at a predetermined time $T$, the experiment concludes, and a feature is launched if the estimated cumulative lift $\hat{\theta}_T$ exceeds the threshold $\theta^\star=z_\alpha\,\mathrm{SE}_T$, where $\mathrm{SE}_T$ is the standard error of $\hat{\theta}_T$ and $z_\alpha$ is the $(1-\alpha)$-quantile of the standard normal distribution.

While this approach controls the false positive rate (FPR) at level $\alpha$ and guarantees adequate power at the design effect, the fixed-horizon assumption becomes inefficient as organizations scale. Constrained testing resources force product teams to wait until $T$ even when early data strongly indicates an obvious success or failure~\cite{gupta2019top}. Consequently, experimenters frequently monitor results continuously and stop tests early (``peeking''), a practice that inflates the false positive rate and invalidates traditional statistical guarantees~\cite{Fiez2024Anduril}.

Continuous monitoring is often formalized through always-valid inference (AVI)~\cite{johari2022always,maharaj2023anytime}, which bounds the error rate uniformly across an infinite time horizon. However, the open-ended nature of AVI conflicts with the practical requirements of product teams who rely on a definitive maximum horizon $T$ to force a launch decision. When a strict deadline $T$ is imposed on an AVI framework, the design becomes overly conservative, allocating error budget to infinite horizons that will never be realized~\cite{ter2024anytime}. Group Sequential Design (GSD) offers a more efficient alternative by calibrating decision boundaries across a finite set of interim looks~\cite{pocock1977group, obrien1979multiple, jennison2000group}.

Despite its theoretical advantages, GSD is rarely implemented in large-scale platform experimentation due to several interconnected operational barriers. First, practitioners are reluctant to adopt early stopping rules if doing so alters the terminal decision criteria established by the platform. Second, standard sequential boundaries assume independent increments, whereas platform metrics consistently exhibit strong daily autocorrelation. Third, the frequentist power of a sequential design cannot be directly verified on online A/B tests because the true effect size distribution is latent. Fourth, while traditional sequential methods optimize for sample size, platform engineering requires optimizing for expected calendar duration to maximize feature velocity. Lastly, the guarantees of a sequential design hold only if its recommendations are executed immediately, whereas in practice, experiment owners will often wait to assess guardrails before acting.

This paper develops a group sequential testing framework tailored to the operational realities of large-scale experimentation. Section~\ref{sec:method} presents an end-to-end methodology for policy construction, detailing how we preserve the terminal decision rule, correct for longitudinal autocorrelation, estimate unobserved operating characteristics, and optimize for expected duration. Section~\ref{sec:results} compares our proposed policy against well-known GSD baselines in both simulation and empirical applications. Section~\ref{sec:practice} discusses enforcing the stopping rule with autonomous agents and correcting the resulting selection bias. Section~\ref{sec:conclusion} outlines future directions for adaptive sampling.

\section{Methodological Framework}
\label{sec:method}

\begin{figure}[htb]
\centering
\includegraphics[width=\textwidth]{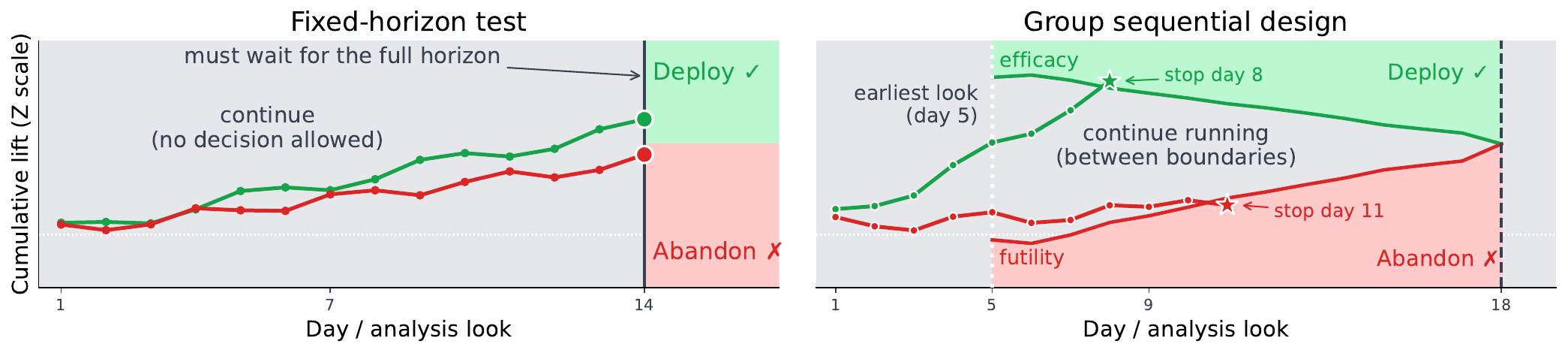}
\caption{Decision boundaries under the fixed-horizon paradigm (left) and the proposed GSD policy (right). The sequential boundaries are constructed to converge at the fixed-horizon terminal threshold.}
\label{fig:boundaries}
\end{figure}

\paragraph{Policy Formulation and Terminal Agreement.} Let $\hat{\theta}_t$ denote the estimated cumulative lift conditional on all observations up to day $t$~\cite{zaidi2025bayesian}. A group sequential policy evaluates $\hat{\theta}_t$ at distinct intervals against a lower futility boundary $\tlo(t)$ and an upper efficacy boundary $\thi(t)$ (Figure~\ref{fig:boundaries}). The decision rule is defined as:
\begin{equation*}
    \textbf{launch} \iff \hat{\theta}_t > \thi(t), \quad \textbf{abandon} \iff \hat{\theta}_t < \tlo(t), \quad \textit{else \textbf{continue}}.
\end{equation*}

Our objective is to design a policy that minimizes the expected duration of experiments while adhering to a target FPR and Power. We require that the terminal rule of the sequential policy matches the platform's established fixed-horizon rule. This yields the following constrained optimization problem:
\begin{equation*}
    (\tlo, \thi) \;=\; \arg\min \;\; \underbrace{\mathbb{E}[\tau]}_{\text{primary objective}}
    \quad \text{s.t.} \quad
    \underbrace{\tlo(T) = \thi(T) = \theta^\star}_{\text{terminal agreement}},
    \quad
    \underbrace{\text{FPR} \le \alpha, \; \text{Power} \ge 1-\beta}_{\text{error constraints}}
\end{equation*}
where $\tau\le T$ is the stopping time.

Standard group sequential designs (e.g., Pocock~\cite{pocock1977group}, O'Brien--Fleming~\cite{obrien1979multiple}) do not satisfy the terminal agreement constraint, as they raise the terminal value to allocate error budget to earlier interim analyses. A design that leaves the terminal rule untouched is comparatively easy to adopt: if no boundary is crossed, the experiment runs to its horizon and is evaluated by the exact test the platform would have applied anyway, so there is no second framework to learn and no final verdict to re-justify. The one downside is the horizon itself, which we treat as a design variable; an experiment that never crosses a boundary may run somewhat longer than the platform's default, and we quantify that trade-off in Section~\ref{sec:results}.

To achieve this, we formulate the boundaries using the Predictive Probability of Success (PPoS)~\cite{zaidi2025bayesian}---the Bayesian predictive probability, given the data observed through day $t$, that the cumulative lift at horizon $T$ will exceed the terminal threshold:
\begin{equation*}
\mathrm{PPoS}_t = \Pr\big(\hat{\theta}_T > \theta^\star \mid \text{data through day } t\big).
\end{equation*}
The policy is parameterized by probability thresholds $(X, Y)$, such that the trial stops for efficacy if $\mathrm{PPoS}_t>X$ and for futility if $\mathrm{PPoS}_t<Y$. At terminal horizon $t=T$, data is fully observed, and $\mathrm{PPoS}_T$ reduces to the indicator function $\ind\{\hat{\theta}_T>\theta^\star\}$. Consequently, this formulation guarantees that the terminal decision aligns with the standard level-$\alpha$ test $\hat{\theta}_T>\theta^\star$. This preserves the decision \emph{rule} rather than each individual decision: experiments stopped early are decided by the interim boundaries, and the selected horizon $T^\star$ may differ from the platform's default.

\paragraph{Deriving Boundaries under Autocorrelation.} Under the standard assumption of independent increments, let $\theta\sim\mathcal{N}(m_0,s_0^2)$ be the prior on the lift, $\mathrm{SE}_t=\sigma/\sqrt{t}$ the standard error after $t$ days (with $\sigma$ the standard deviation of a daily lift increment), and $\kappa(t)=\frac{s_0^2+\mathrm{SE}_T^2}{s_0^2+\mathrm{SE}_t^2}\in(0,1]$ the shrinkage factor of the predictive mean $\mathbb{E}[\hat{\theta}_T\mid\hat{\theta}_t]=m_0+\kappa(t)(\hat{\theta}_t-m_0)$. The probability thresholds map to closed-form boundaries on the lift scale:
\begin{equation*}
\tlo(t) = \frac{\theta^\star-(1-\kappa(t))m_0}{\kappa(t)} -\frac{\Phi^{-1}(1-Y)}{\sqrt{\kappa(t)}}\sqrt{\mathrm{SE}_t^2-\mathrm{SE}_T^2}, \qquad
\thi(t) = \frac{\theta^\star-(1-\kappa(t))m_0}{\kappa(t)} +\frac{\Phi^{-1}(X)}{\sqrt{\kappa(t)}}\sqrt{\mathrm{SE}_t^2-\mathrm{SE}_T^2}.
\end{equation*}

However, longitudinal panel data violates the independent-increments assumption. Because the same users contribute observations on consecutive days, each additional day reduces the standard error less than a day of fresh users would. Treating dependent increments as independent deflates the standard error, yielding permissive boundaries that inflate the FPR and overstate power. To correct this, we estimate the empirical correlation matrix $R$ of the daily increments from historical A/A tests, pooling non-overlapping windows of daily z-scores across experiments and averaging correlations by lag, and replace the unadjusted time $t$ with an effective sample size formulation:
\begin{equation*}
n_{\mathrm{eff}}(t) = \frac{t^2}{\sum_{i,j\leq t}R_{ij}}, \qquad \mathrm{SE}_t^{\mathrm{eff}}=\frac{\sigma}{\sqrt{n_{\mathrm{eff}}(t)}}.
\end{equation*}
This reduces to $n_{\mathrm{eff}}(t)=t$ under independence ($R=I$), and satisfies $n_{\mathrm{eff}}(t)\le t$ whenever the average correlation among the first $t$ days is non-negative. Substituting $\mathrm{SE}_t^{\mathrm{eff}}$ into the boundary equations at every look, including the terminal $\mathrm{SE}_T^{\mathrm{eff}}$ in $\kappa(t)$, in the square-root term, and in the threshold $\theta^\star=z_\alpha\,\mathrm{SE}_T^{\mathrm{eff}}$, yields the adjusted, autocorrelation-aware boundaries $(\tlo^{\mathrm{eff}}(t),\thi^{\mathrm{eff}}(t))$.

\paragraph{Evaluating Unobserved Operating Characteristics.} Verifying the FPR and power of a sequential procedure on empirical data poses an estimation problem: the point-null FPR requires a strict null hypothesis, and power is a function of a latent distribution of true effect sizes. Following Zhou et al.~\cite{zhou2023bayesian}, we estimate the average operating characteristics by integrating over the empirical distribution. Using a mixture of A/A tests (known nulls) and A/B tests, we define $L_e=\ind\{\hat{\theta}_{e,14}>\theta^\star\}$ and $D_e=\ind\{\text{GSD launches }e\}$. Let $\hat{p}_R^S$ denote the launch rate of rule $R\in\{L,D\}$ in set $S\in\{A/A,A/B\}$. We define the structural estimators:
\begin{equation*}
\hat{\pi}_1 =\frac{\hat{p}_L^{A/B}-\hat{p}_L^{A/A}}{(1-\beta)-\hat{p}_L^{A/A}} \ \ (\text{Prevalence}), \qquad \hat{\alpha}_D=\hat{p}_D^{A/A} \ \ (\text{FPR}), \qquad \widehat{1-\beta}_D =\frac{\hat{p}_D^{A/B}-(1-\hat{\pi}_1)\hat{\alpha}_D}{\hat{\pi}_1} \ \ (\text{Power}).
\end{equation*}
Because empirical launch rates identify only the product $\pi_1(\bar{P}_L-\alpha_L)$---where $\bar{P}_L$ and $\alpha_L$ are the average power and FPR of the fixed-horizon test---we anchor $\bar{P}_L$ at its nominal value $1-\beta$ (the nominal power at the minimum detectable effect). This anchor only sets the scale: mathematically, $\widehat{1-\beta}_D=\hat{\alpha}_D+\hat{\rho}\,(1-\beta-\hat{p}_L^{A/A})$, where the ratio $\hat{\rho}=\frac{\hat{p}_D^{A/B}-\hat{p}_D^{A/A}}{\hat{p}_L^{A/B}-\hat{p}_L^{A/A}}$ requires no parametric assumptions. Consequently, verifying that $\widehat{1-\beta}_D\ge 1-\beta$ amounts to verifying that the sequential design retains the power of the fixed-horizon test over the unobserved latent effect distribution (i.e., verifying $\hat{\rho}\gtrsim 1$ when $\hat{\alpha}_D\approx\hat{p}_L^{A/A}$).

\paragraph{Optimizing for Expected Duration.} We solve the parameter space $(X, Y, T^\star)$ entirely via simulation. We simulate trajectories utilizing the empirical correlation matrix $R$ under both the null ($\theta=0$) and a point alternative at the MDE. The design problem reduces to a grid search over the stopping boundaries, selecting the $(X,Y,T)$ that minimizes the expected stopping day $\bar{\tau} = (1-\hat{\pi}_1)\bar{\tau}_0 + \hat{\pi}_1\bar{\tau}_1$, subject to $\mathrm{FPR}\leq\alpha$ and $\mathrm{Power}\geq 1-\beta$. Because $\theta^\star=z_\alpha\,\mathrm{SE}_{T^\star}$, the terminal test maintains level $\alpha$ for any selected horizon $T^\star$, so treating the horizon as a design variable does not compromise the terminal guarantee. These optimal parameters are then used in online experiments without further re-tuning.

\section{Empirical Evaluation}
\label{sec:results}

We evaluate the proposed \textsc{PPoS} framework against a standard 14-day fixed-horizon design~\cite{Radwan2024Eval, Malmir2026ImpressionShare, Malmir2026BeyondIPS}, alongside O'Brien--Fleming (OBF)~\cite{obrien1979multiple} and Pocock~\cite{pocock1977group} sequential boundaries, and a likelihood-ratio (LR) oracle bound~\cite{jennison2000group}. A comparison against always-valid inference is deferred to an extended version of this work: calibrating an open-ended process to match a specific FPR at a fixed horizon involves design choices that materially affect the resulting conservatism. All sequential policies share the same design parameters: they are calibrated to an FPR $\le 0.10$ and power $\ge 0.80$ at the MDE, perform interim analyses daily from day 5 onward, and select their horizon from the same candidate grid $T\in\{10,\dots,21\}$. Calibration uses $\approx 10^5$ simulated trajectories drawn under the empirical correlation matrix $R$. For \textsc{PPoS}, we assign an uninformative prior ($s_0^2\to\infty$), which yields $\kappa(t)=1$ and collapses the boundaries to a frequentist object while retaining the Bayesian infrastructure for future integration of historical priors. More generally, $\mathrm{PPoS}_t$ is the posterior expectation of the conditional power $\Pr(\hat{\theta}_T>\theta^\star\mid\hat{\theta}_t,\theta)$, averaged over the remaining posterior uncertainty in $\theta$ rather than evaluated at a single plug-in value.

Empirical validation is performed on approximately 2000 online A/A and A/B experiments tracking an internal supply metric. The A/A and A/B experiments are drawn from the same population and run under the same traffic allocation, duration, and logging configuration, differing only in whether a treatment is applied; the A/A tests are therefore a valid reference both for the point-null FPR $\hat{\alpha}_D$ and for the correlation matrix $R$. Boundaries established in simulation are applied directly to the online experiments without re-tuning. Online $95\%$ confidence intervals are percentile bootstrap intervals ($1000$ replicates) that resample A/A and A/B experiments separately and recompute all estimators, including $\hat{\pi}_1$. Calibration and validation share $\hat{\pi}_1$ and $R$, but the resulting leakage is limited: $\hat{\pi}_1$ weights only the objective and not the error constraints, and $R$ uses no launch outcomes.

\begin{figure}[H]
\centering
\includegraphics[width=\textwidth]{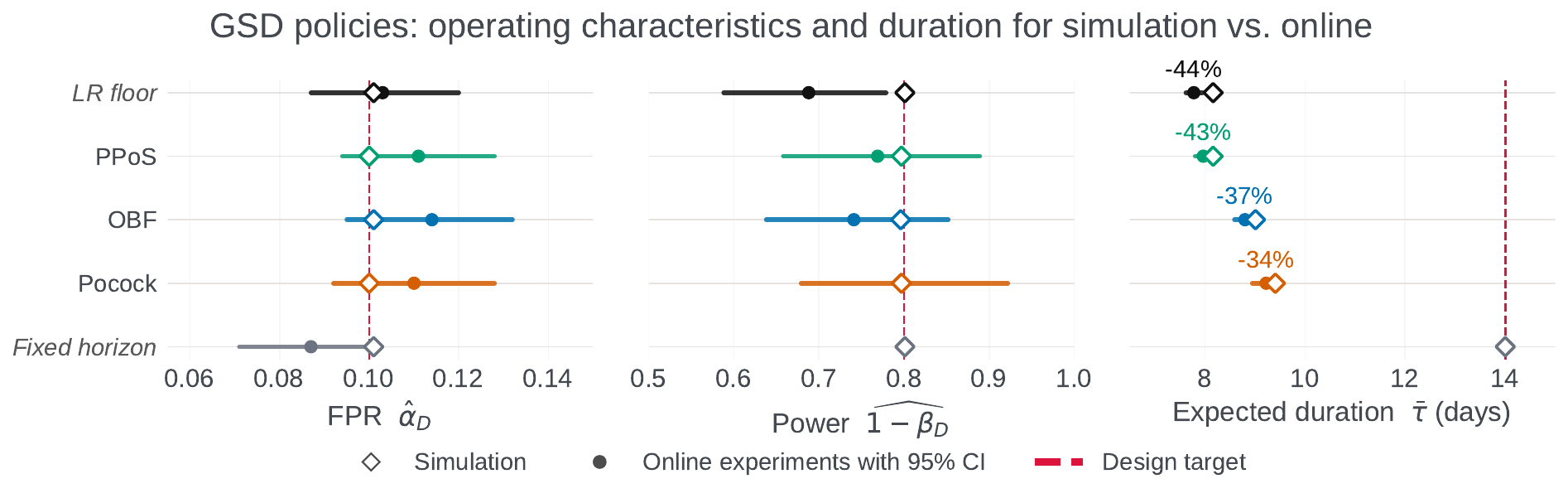}
\caption{FPR, average power, and expected duration across evaluated designs. Simulated properties are denoted by diamonds; empirical properties evaluated on online experiments are denoted by dots with $95\%$ confidence intervals. The online power of the fixed-horizon design is the anchor of the power estimator and is therefore not shown.}
\label{fig:results}
\end{figure}

Simulation results (Figure~\ref{fig:results}, diamonds) indicate that the \textsc{PPoS} framework attains the efficiency floor defined by the LR oracle bound, achieving an expected duration of $\bar{\tau} = 8.16$ days, a $41.7\%$ reduction relative to the 14-day baseline. This is expected rather than coincidental: at each look, $\mathrm{PPoS}_t$ is a monotone function of the likelihood ratio, so the policy lies within the class of threshold rules over which the LR floor is optimal, and its two thresholds $(X,Y)$ suffice to attain it. The grid search selects an optimal horizon of $T^\star=17$; under a constraint forcing $T=14$, the efficiency improvements largely persist, yielding $\bar{\tau}=9.17$ days (a $34.5\%$ reduction). At its selected horizon, the framework outperforms both OBF ($\bar{\tau}=9.01$ days, $T^\star=17$) and Pocock ($\bar{\tau}=9.41$ days, $T^\star=21$); because OBF selects the same horizon, that comparison is horizon-matched and the remaining gap reflects boundary shape alone.

Applying the framework to empirical data confirms that the structural estimators remain calibrated (Figure~\ref{fig:results}, dots). For \textsc{PPoS}, OBF, and Pocock, the $95\%$ bootstrap confidence intervals for both the empirical FPR $\hat{\alpha}_D$ and the average power $\widehat{1-\beta}_D$ contain their nominal targets. The online power of the LR floor, by contrast, falls below $0.80$---an expected outcome, since that bound is optimal only for the simulated two-point model in which every true effect equals the MDE. In the empirical application, the \textsc{PPoS} policy yielded an expected duration of $\bar{\tau} = 7.97$ days $[7.81, 8.15]$. Operationally, this $43.1\%$ reduction in average experiment runtime translates to a $76\%$ increase in aggregate platform testing capacity, assuming constant experiment generation and a fixed pool of testing slots released when an experiment stops rather than held to its horizon.

\section{GSD In Practice and Post-Selection Inference}
\label{sec:practice}

The validity of the sequential estimators relies on the assumption that the trial terminates at the stopping time $\tau$. In practice, an operational delay often occurs between a policy triggering and a human experimenter taking action. This hesitation is frequently driven by a desire to ``further validate'' an early outcome via additional data---a form of \emph{anti-peeking}---or to wait for guardrail metrics to stabilize.

This behavioral delay confounds the design, and the two boundaries fail in different ways. Ignoring a futility crossing and continuing data collection grants the procedure subsequent, un-modeled opportunities to cross $\thi$ later, invalidating FPR control. Conversely, delaying a launch after an efficacy crossing is conservative for the FPR, but it forfeits the calendar time the policy was explicitly optimized to save and leaves the realized stopping distribution different from the one against which the design was calibrated. To preserve the statistical properties, execution of the stopping rule should be enforced programmatically by an automated agent. Unlike traditional automation, agentic integrations can provide detailed, human-readable reasoning for the termination, fit into existing summarization workflows, and defer only when a pre-specified data-quality check fails---making the exception part of the calibrated policy rather than a discretionary override.

Furthermore, conditioning the analysis on the stopping time $\tau$ introduces selection bias into the point estimate, a phenomenon generally known as the winner's curse~\cite{tsiatis1984exact, emerson1990parameter}. Because the PPoS framework solely defines the stopping rule, effect estimation requires secondary adjustment. To provide an unbiased post-selection estimate, the terminal values should be evaluated via a conditional estimator that explicitly accounts for the boundary-crossing distribution~\cite{andrews2024inference}, or smoothed using Bayesian shrinkage under a conservative prior centered at zero~\cite{mudd2026breaking,zhou2023bayesian}.

\section{Conclusion}
\label{sec:conclusion}

We propose a group sequential testing framework that addresses the operational constraints of online experimentation. By formulating boundaries via the Predictive Probability of Success and correcting for longitudinal autocorrelation via an effective sample size adjustment, the policy preserves the fixed-horizon terminal rule while controlling error rates. Evaluation on empirical distributions demonstrates that the framework reduces expected duration by $43.1\%$, increasing effective platform capacity by $76\%$, and matches the theoretical oracle bound in simulation. Delegating execution to an agent is what would keep these guarantees valid in practice, and we have sketched the design of such a mechanism. Future extensions of this framework will investigate integrating sequential boundaries with adaptive sampling mechanisms, such as interim sample size re-estimation~\cite{gsponer2014} and dynamic allocation rules~\cite{Balagopalan2026FC2FB}.

\bibliographystyle{unsrt}
\bibliography{refs}

\end{document}